\documentclass[jpco,reprint,twocolumn,showpacs,nofootinbib,unsortedaddress]{revtex4}
\usepackage{amsfonts}
\usepackage{amsmath}
\usepackage{amssymb}
\usepackage{upgreek}
\usepackage{bm}
\usepackage{dcolumn}
\usepackage{epsfig}
\usepackage{graphicx}
\usepackage{graphics}
\usepackage{placeins}
\usepackage{subfigure}
\newcommand{\R}{\mathbf{R}}

\newcommand{\Q}{\mathbf{Q}}
\newcommand{\uu}{\mathbf{u}}
\usepackage{color}

\usepackage{graphicx, subfigure}
\usepackage{float}
\usepackage{amsmath}
\usepackage{acronym}
\usepackage{multirow}
\usepackage{amsfonts}
\usepackage{ulem}

\newcommand{\beq}{\begin{equation}}
\newcommand{\eeq}{\end{equation}}
\newcommand{\bea}{\begin{eqnarray}} 
\newcommand{\eea}{\end{eqnarray}}
\newcommand{\ba}{\begin{array}}
\newcommand{\ea}{\end{array}}

\begin{document} 

\title{Proving the short-wavelength approximation in Pulsar Timing Array gravitational-wave background searches}

\pacs{%
04.80.Nn, 
04.25.dg, 
97.60.Gb, 
04.30.Tv 
}
\author{Chiara M.~F.~Mingarelli}
\affiliation{Center for Computational Astrophysics, Flatiron Institute, 162 5th Ave, New York, NY 10010 }
\email{cmingarelli@flatironinstitute.org}

\author{Angelo B.~Mingarelli}
\affiliation{School of Mathematics and Statistics,
Carleton University, Ottawa, Ontario, Canada, K1S\, 5B6}
\email{angelo@math.carleton.ca}

\keywords{asymptotics, gravitational waves, stationary phase, black holes, pulsar timing}

\begin{abstract}
A low-frequency gravitational-wave background (GWB) from the cosmic merger history of supermassive black holes is expected to be detected in the next few years by pulsar timing arrays. A GWB induces distinctive correlations in the pulsar residuals --- the expected arrival time of the pulse less its actual arrival time. Simplifying assumptions are made in order to write an analytic expression for this correlation function, called the Hellings and Downs curve for an isotropic GWB, which depends on the angular separation of the pulsar pairs, the gravitational-wave frequency considered, and the distance to the pulsars. This is called the short-wavelength approximation, which we prove here rigorously and analytically for the first time.
\end{abstract}

\maketitle
\numberwithin{equation}{section}
\newtheorem{theorem}{Theorem}[section]
\newtheorem{lemma}{Lemma}[section]
\newtheorem{corollary}[theorem]{Corollary}
\newtheorem{example}[theorem]{Example}

\section*{Introduction}
Gravitational waves (GWs) are ripples in the fabric of space-time, originating from some of the most violent
events in the Universe, including the mergers of supermassive black holes. High frequency GWs from the merger of stellar-mass black holes were first detected by the 
Laser Interferometer Gravitational-wave Observatory (LIGO) in September 2015 \cite{AbbottEtAl:2016}, hailing the dawn of gravitational-wave astronomy. However, LIGO can only
detect high frequency GWs, in the 100 - 1000 Hz range. Similarly to electromagnetic radiation, different GW detectors are needed to probe different GW frequencies.
Currently there are plans to launch a space-based GW detector in 2034 -- the Laser Interferometer Space Antenna (LISA) \cite{LISA:2017} -- which will probe the millihertz
GW frequency regime, thought to be populated primarily by merging supermassive black holes (SMBHs) in the $10^5-10^6~M_\odot$ range. 
At the very low-frequency end of the GW spectrum, one expects to find nanohertz GWs from very massive inspiraling SMBHs, in the $10^8-10^9~M_\odot$ range. These
can be detected by timing millisecond pulsars, called a Pulsar Timing Array (PTA) \cite{s78,det79,hd83, bcr83}. Millisecond pulsars are excellent clocks, and delays or advances
in their arrival times -- inducing a timing residual -- could signal the presence of GWs. PTA experiments are very active, and have been taking data for over a decade \cite{abb+18a,desvignes+:2016, ShannonEtAl:2015}. With a PTA, one can detect not only
GWs from inspiraling SMBH binaries (SMBHBs), see e.g. \cite{bps+15, mls+2017}, but the GW background (GWB) from the cosmic merger history of SMBHBs \cite{abb+18b, ltm+15,VerbiestEtAl:2016}. This GWB is expected to be detected in the next few years \cite{sejr13, tve+16},
with the details depending on the underlying astrophysics of the SMBH mergers. More details on PTAs can be found in recent review articles, e.g. \cite{t18, h17, l15, sbs15}, and an outline GW astrophysics covering nanohertz to kilohertz frequencies can be found in \cite{bcn+18}.

Indeed, a rigorous exploration and examination of the tools which will be used to make the first detection of a GWB is crucial. An isotropic GWB will 
induce characteristic correlations in the pulsar timing residuals. By cross-correlating these residuals, one expects to see a characteristic correlation called the Hellings and Downs curve \cite{hd83}.
Deviations from an isotropic GWB can be induced by nearby and/or particularly loud SMBHBs, inducing anisotropy in the GWB. Anisotropic GWBs will induce different correlations patterns, and have been explored by \cite{msmv13, tg13, ms14,grt+14, cvh14}. 

Here we prove analytically, and for the first time, that the Hellings and Downs curve can be extracted from the cross-correlated pulsar residuals, without making assumptions that the pulsars are all at the same distance $L$ from the Earth. Part of this proof is a  consequence of the application of the Riemann-Lebesgue Lemma and the Lebesgue Dominated Convergence Theorem -- well-known in the mathematics community, but somewhat obscure in the field of GWs. We emphasize that no previous work has been able to do this analytically, though computer-aided integration has been used to verify one's intuition numerically.

\section{The characteristic strain}
The International PTA (IPTA) published combined data on 49 millisecond pulsars in their first data release~\cite{VerbiestEtAl:2016}. These millisecond pulsars are the most stable natural astrophysical clocks known~\cite{taylor91}, and are regularly monitored by 8 radio telescopes: 5 in Europe~\cite{desvignes+:2016}, 2 in North America~\cite{abb+18a} and one in Australia~\cite{ShannonEtAl:2015}. PTAs take advantage of the precise arrival times of millisecond pulsars to enable GW detection. 

The GWB is described in terms of its characteristic strain, $h_c(f)$, with amplitude $A$ at a reference frequency of $1/$yr (e.g. \cite{p01}):
\beq
h_c(f) = A \left( \frac{f}{\mathrm{yr}^{-1}}\right)^{-2/3}\, .
\eeq 

The current upper limits on $A$ are difficult to compare, since it was recently discovered that errors in planetary masses and positions (called the solar system ephemeris model) can directly affect the limit on $A$~\cite{abb+18b, thk+16}, and in some cases mimic a GWB signal. 

\begin{figure*}
		\centering
		\includegraphics[width=4in]{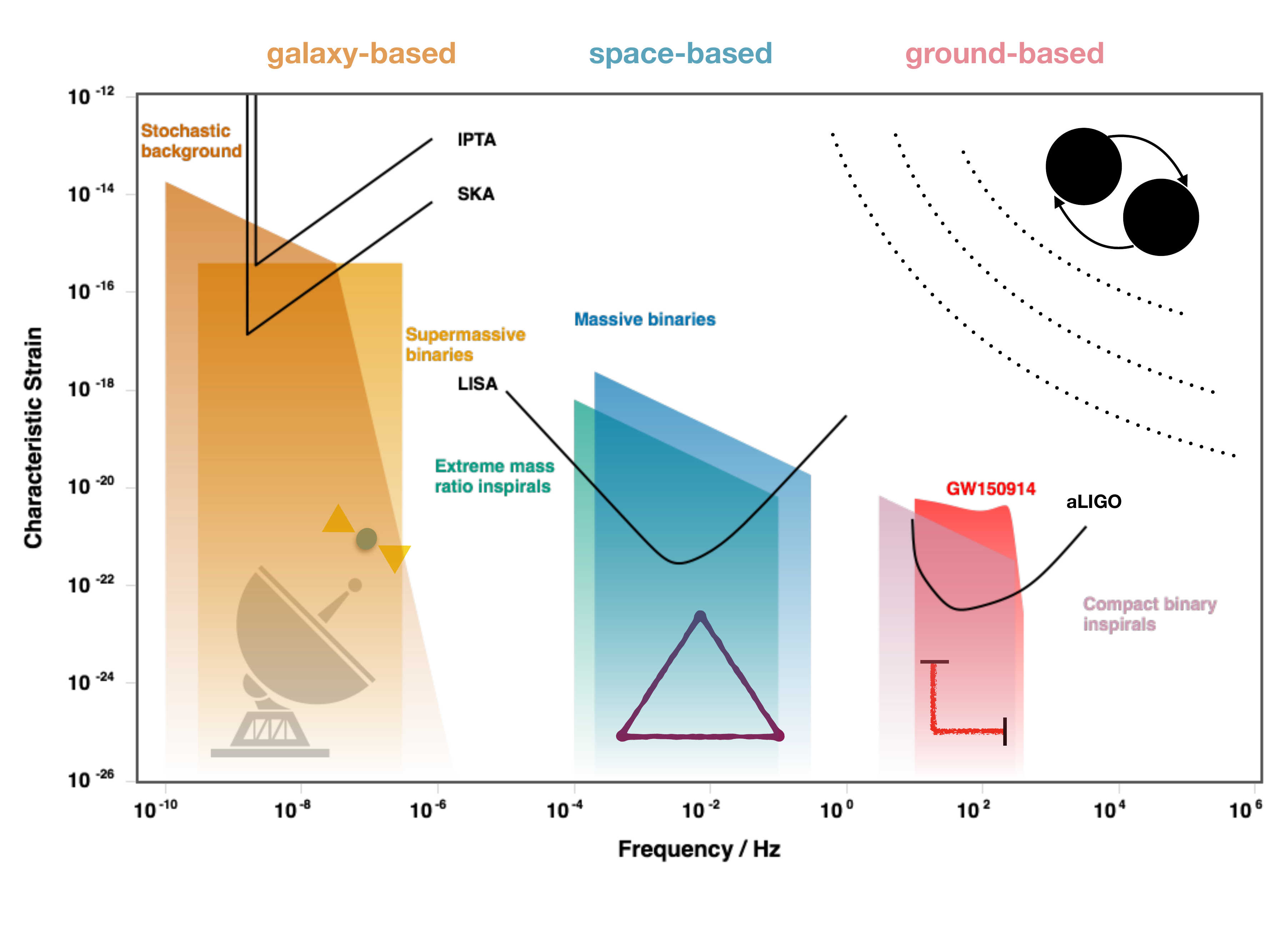}
		\caption[The Gravitational-Wave Landscape]{The spectrum of gravitational radiation from low-frequency to high-frequency. At very low frequencies pulsar timing arrays can detect both the GWB from supermassive black hole binaries, in the $10^8-10^{10}~M_\odot$ range, as well as radiation from individual binary sources which are sufficiently strong. For the IPTA sensitivity we assume 20 pulsars with 100 ns timing precision with a 15 year dataset, and for the SKA we assume 100 pulsars timed for 20 years with 30~ns timing precision. Both estimates assume 14-day observation cadence. The PTA spans the size of the galaxy, and is therefore a ``galaxy-based'' GW detector. LISA is a space-based GW detector scheduled to launch in 2034~\cite{LISA:2017}. aLIGO a ground-based high-frequency GW detector, and is currently the only detector to directly detect GWs from compact binary coalescences, which currently include binary black holes and binary neutron star mergers~\cite{AbbottEtAl:2016, Abbott+2017}. Note that these GW detectors are all complementary, and that LIGO cannot, for example, detect GWs from supermassive black hole binaries, just as PTAs cannot detect high-frequency GWs from merging stellar-mass black holes. A review of current and future GW detectors across the spectrum is available in \cite{bcn+18}.} 
		\label{fig:gwspectrum}
\end{figure*}

While the current upper limit on $A$ from NANOGrav can take this into account, and limit $A < 1.35 \times 10^{-15}$, other PTAs have not yet published updates to their limits. Projections the characteristic strain accessible with future IPTA and Square Kilometer Array (SKA)~\cite{ska:09, JanssenEtAl:2015, mcb15} detectors are shown in Figure~\ref{fig:gwspectrum}.

The observed residuals due to the presence of a GWB with characteristic strain $h_c(f)$ is described by the cross-power spectral density of pulsar 1 and pulsar 2 by
\beq
S_{1,2}(f) = \frac{\Gamma_{1,2}(fL_1, fL_2, \zeta) h^2_c(f)}{12 \pi^2 f^3}\, ,
\eeq
see e.g. \cite{jhs+06}, where $\Gamma_{1,2}$ is the so-called overlap reduction function, which describes the GWB-induced correlation signature in the pulsar residuals. This is a function of the frequency of the GWB, the distance to the pulsars $L_{1,2}$, and the angular separation of the pulsars, $\zeta$. PTA geometry is explored in detail in Figure \ref{fig:geometry}. For an isotropic GWB, this is called the Hellings and Downs curve~\cite{hd83}, and for anisotropic GWBs see \cite{msmv13, tg13, ms14, grt+14}.

\section{The Hellings and Downs curve}
\begin{figure}
		\centering
		\includegraphics[width=3.5in]{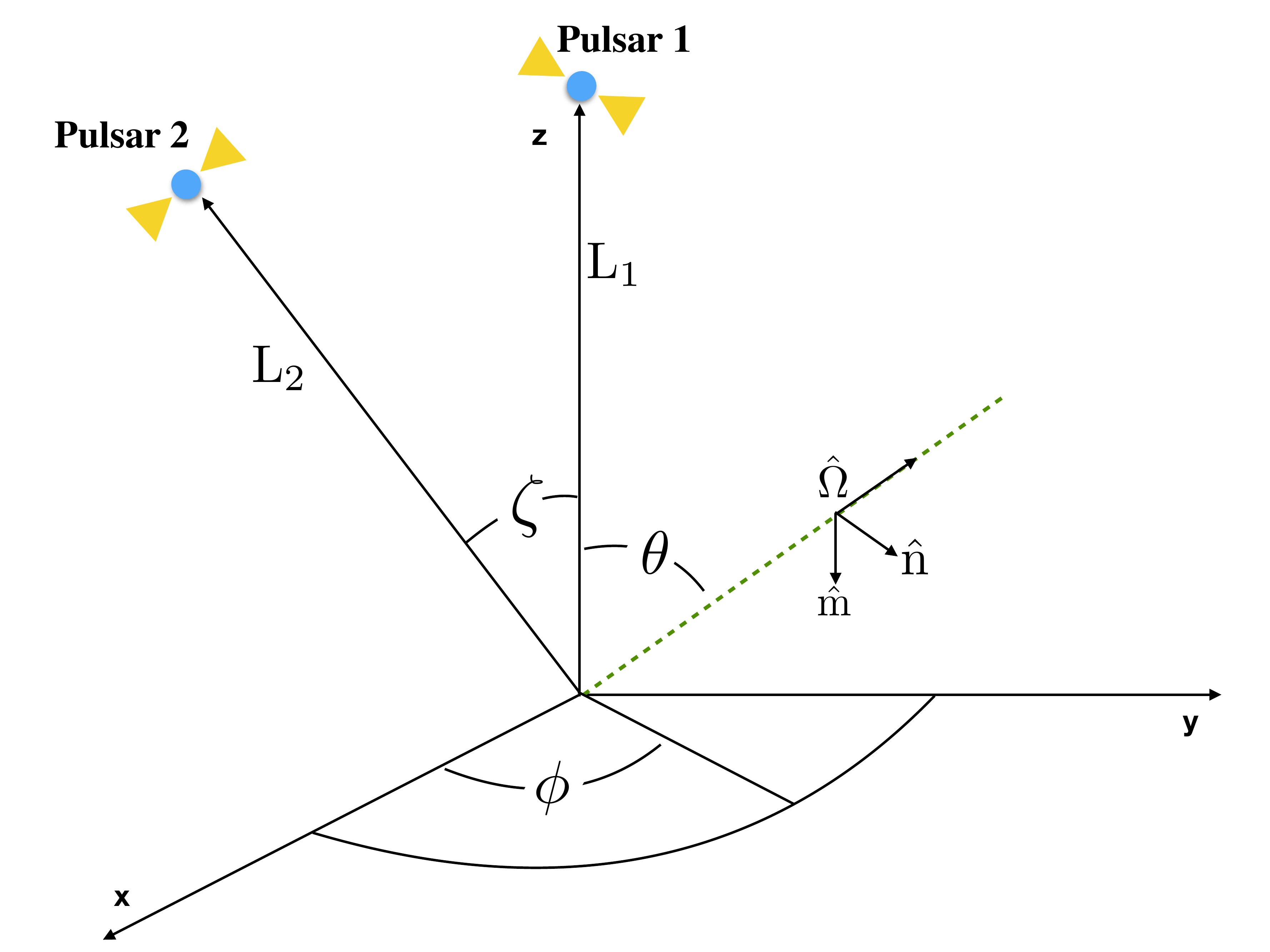}
		\caption[Typical pulsar timing array geometry]{Pulsar $1$ is on the $z$-axis at a distance $L_1$ from the origin, and Pulsar $2$ is in the $x$-$z$ plane at a distance $L_2$ from the origin making an angle $\zeta$ with Pulsar $1$. Here $\hat\Omega$ is the direction of GW propagation, with principal axes $\hat m$ and $\hat n$, such that  $\hat m \times \hat n =\hat\Omega$. The angles $\theta \in [0,\pi]$ and $\phi \in [0,2\pi]$ are the polar and azimuthal angles, respectively.}
		\label{fig:geometry}
\end{figure}

In analogy with \cite{msmv13, ms14}, we present an overview of how one arrives to the Hellings and Downs curve. A source of GWs in direction $-\hat\Omega$, see Figure \ref{fig:geometry},
generates a metric perturbation $h_{ij}(t,\hat{\Omega})$, which we describe as a plane wave:

\beq
\label{eq:plane}
h_{ij}(t, \vec{x}) = \sum_A \int_{-\infty}^\infty df \int_{S^2} d\hat\Omega h_A(f,\hat\Omega)e^{i2\pi f(t-\hat\Omega\cdot\vec{x})}e^A_{ij}(\hat\Omega)\, .
\eeq
This  can be decomposed over two 
polarization tensors $e^A_{ij}(\hat\Omega)$, and two independent polarization amplitudes $h_A(t,\hat\Omega)$~\cite{mtw, ar99}:
\beq
\label{eq:metHij}
h_{ij}(t,\hat\Omega) = e^+_{ij}(\hat\Omega)h_+(t,\hat\Omega)+e^\times_{ij}(\hat\Omega)h_\times(t,\hat\Omega)\, .
\eeq
We note that General Relativity predicts only two independent polarizations,
plus $+$, and cross $\times$, while other theories predict additional polarizations, such as breathing modes \cite{ljp08, ss12, grt15}. He we restrict ourselves to the well-known tensor polarizations, $A=+,\times$.

The $e^A_{ij}(\hat\Omega)$ polarization tensors are uniquely defined by specifying $\hat m$ and $\hat n$ -- the GW principal axes, illustrated in Figure \ref{fig:geometry}:

\beq
e^+_{ij}(\hat\Omega) = \hat m_{i}\hat m_{j} - \hat n_{i}\hat n_{j} \, , \hspace{1cm} e^\times_{ij}(\hat\Omega) = \hat m_{i}\hat n_{j} + \hat n_{i}\hat m_{j}\, .
\eeq
For a stationary, Gaussian, and unpolarized GWB, the polarization amplitudes satisfy (see e.g. \cite{FLR_09}):
\beq
\langle h^*_A(f,\hat\Omega)h_{A'}(f',\hat\Omega')\rangle = \delta^2(\hat\Omega, \hat\Omega')\delta_{AA'}\delta({f-f'})H(f).
\eeq

The metric perturbation will change the proper distance between the Earth and the pulsars, inducing an advance or delay in the pulsar pulse's arrival time at the Earth. Consider for example a millisecond pulsar with frequency $\nu_0$ whose location in the sky is described by $\hat{p}$, at a distance $L$ from the Earth. The metric perturbation affects the frequency of the radio pulses, $\nu$, received at the radio telescope. This frequency shift is given by

\beq
z(t,\hat\Omega)  \equiv  \frac{\nu(t) - \nu_0}{\nu_0} = \frac{1}{2} \frac{\hat p^i\hat p^j}{1+\hat \Omega \cdot \hat p} \Delta h_{ij}(t,\hat{\Omega})\,,
\label{e:z}
\eeq

where
\beq
\Delta h_{ij}(t,\hat\Omega) \equiv h_{ij}(t_e,\hat{\Omega}) - h_{ij}(t_p,\hat{\Omega}) 
\label{e:deltah}
\eeq
is the difference between the GW-induced metric perturbation at the Earth $h_{ij}(t_e,\hat{\Omega})$, the {Earth term}, with coordinates $(t_e,\vec{x}_e)$, and at the pulsar $h_{ij}(t_p,\hat{\Omega})$, the pulsar term, with coordinates $(t_p,\vec{x}_p)$:
\begin{align}
t_p & = t_e - L\, ,  \quad \quad \vec{x}_p = L \hat p  \, , \quad \quad \vec{x}_e = 0\,.
\end{align}
The indices ``e'' and ``p'' refer to the Earth and the pulsar, however, it is standard write $t_e=t$, see e.g. \cite{abc+09, msmv13, ms14, mtg15}.

We can now write \eqref{e:deltah}, using \eqref{eq:plane} and \eqref{eq:metHij} as 
\bea
\Delta h_{ij}(t,\hat \Omega) = &&
\sum_A \int_{-\infty}^\infty df e_{ij}^A(\hat \Omega) \ h_A(f,\hat \Omega)\times \nonumber\\
&&e^{i 2 \pi f t} \left[1 - e^{-i 2 \pi f L (1+ \hat \Omega \cdot \hat p)}\right]\, . 
\label{e:deltah1}
\eea
The fractional frequency shift, $z(t)$, produced by a stochastic GWB is simply given by integrating Eq.~(\ref{e:z}) over all directions. Using \eqref{eq:plane} and (\ref{e:deltah1}), we obtain:
\bea 
\label{eq:freqShift}
z(t)  &=& \int d\hat\Omega\, z(t,\hat\Omega) \\
& = & \sum_A \int_{-\infty}^\infty df \int_{S^2}d\hat\Omega F^A(\hat\Omega) h_A(f,\hat \Omega) e^{i 2 \pi f t}\times \nonumber \\ 
&& \left[1 -  e^{-i 2 \pi f L (1+ \hat \Omega \cdot \hat p)} \right],
\eea
where $F^A(\hat\Omega)$ are the antenna beam patterns for each polarization $A$, which we write as
\beq
F^A(\hat\Omega) =\left[\frac{1}{2} \frac{\hat p^i\hat p^j}{1+\hat \Omega \cdot \hat p} \ e_{ij}^A(\hat \Omega)\right].
\label{e:F+Fx}
\eeq

Searches for the GWB rely on looking for correlations induced by GWs in the timing residuals of pulsar pairs. Indeed, the observed quantity in PTA experiments is the timing residual $r(t)$, which is simply the integral of Eq.~(\ref{eq:freqShift}) in time:
\beq
r(t) = \int^t dt' z(t')\,. \\
\label{e:r}
\eeq
The expected value of the correlation between a residual from pulsar $1$ at time $t_j$, with that from a different pulsar, say pulsar $2$ at time $t_k$, depends on terms of the form:

\beq
\langle r_1^*(t_j) r_2(t_k) \rangle = 
\left\langle \int^{t_j} dt' \int^{t_k} dt'' z_1^*(t') z_2(t'') \right\rangle \, ,
\eeq

\begin{widetext}
\beq
\langle r_1^*(t_j) r_2(t_k) \rangle =  \int^{t_j} dt' \int^{t_k} dt'' \int_{-\infty}^{+\infty} df e^{- i 2\pi f(t' - t'')} H(f)\,\Gamma(fL_1, fL_2, \zeta),
\label{e:corr}
\eeq
\end{widetext}
where $H(f)$ contains the information of the spectrum of radiation. In analogy with~\cite{ar99, msmv13, tg13}, we define the quantity above that depends on the angular separation of the pulsars, $\zeta$, their distances from the Earth, $L_1, L_2$, and the GW frequency $f$, as the \textit{overlap reduction function}

\beq\label{eq:HDsol}
\Gamma(fL_1, fL_2, \zeta) \equiv \int d\hat\Omega \, \kappa(f,\hat\Omega) \left[\sum_A F_1^A(\hat\Omega) F_2^A(\hat\Omega)\right],
\eeq
where

\beq
\kappa(fL_{1,2},\hat\Omega) \equiv \left[ 1 - e^{i 2 \pi f L_1 (1+ \hat \Omega \cdot \hat p_1)} \right] \left[ 1 - e^{-i 2 \pi f L_2 (1+ \hat \Omega \cdot \hat p_2)}\right].
\label{eq:kappa}
\eeq

In order to write a closed-form, analytic solution to \eqref{eq:HDsol}, we choose a reference frame where one pulsar is placed along the
$z$-axis and the other in the $x$-$z$ plane as seen in Figure \ref{fig:geometry}. Specifically, we write

\begin{subequations}
	\label{e:coord}
	\begin{align}
	\hat p_1		&=(0,0,1), \\
	\hat p_2	&=(\sin\zeta,0,\cos\zeta),\\
	\hat \Omega 	&=(\sin\theta\cos\phi,\sin\theta\sin\phi,\cos\theta),\\
	\hat m		&=(\sin\phi,-\cos\phi,0),\\
	\hat n		&=(\cos\theta\cos\phi,\cos\theta\sin\phi,-\sin\theta).
	\end{align}
\end{subequations}
We remind the reader that $\hat p_1$ and $\hat p_2$ are the unit vectors pointing to
pulsars $1$ and $2$, respectively, $\hat \Omega$ is the direction of
GW propagation and $\hat m$ and $\hat n$ are the GW principal axes,
see Figure \ref{fig:geometry}. Note that in this reference frame $F_a^\times = 0$ by \eqref{e:F+Fx}, making it a convenient choice. 

For an isotropic GWB one is free to choose whichever coordinate system is most convenient, as was done here. However, one must be more careful when considering 
reference frames which are used to describe pulsar locations in an anisotropic GWB, as was done by \cite{msmv13}.

\vspace{10mm}

\section{Main Results}

\begin{widetext}
We choose the coordinate system defined in Eq. \eqref{e:coord}, and apply it to Eq. \eqref{eq:HDsol}. The result is Eqs. \eqref{eq1} and \eqref{k2}.

{\bf Claim: }\label{thII1}
Let $L_1, L_2, f$, be real positive constants. Then, for each $\zeta \in
(0, \pi]$, as $fL_1 \to \infty$ and $fL_2 \to \infty$,
we have
\label{th1}
\begin{eqnarray}
\label{eq1}
\int_0^\pi \!\! d\theta \!\!
\int_0^{2\pi} \!\!\!d\phi\, K_2(\zeta,\phi,\theta)  \! \left [1- e^ {2\pi {\rm i}  fL_1(1+\cos\theta)} \right]  \! \! \left [1-e^
{-2\pi {\rm i} fL_2(1+\cos\theta\,\cos\zeta+\sin\theta\sin\zeta\cos\phi)} \right]\ 
\! \! \longrightarrow \! \int_0^\pi \!\!\! d\theta \!\!
\int_0^{2\pi} \!\!\! d\phi  K_2(\zeta,\phi,\theta)\,\label{eq2}
\end{eqnarray}
except when $\zeta=0$ and $L_1=L_2$, a case covered in \cite{ms14}. Here, 
\begin{equation}\label{k2}
 K_2(\zeta,\phi,\theta) = \frac{\sin \theta(1-\cos \theta)\ \big (
\sin^2\phi \sin^2\zeta - (\sin\zeta\cos\theta\cos\phi - \sin\theta \cos\zeta)^2\big
)}{1+\cos\theta\cos\zeta+\sin\theta\sin\zeta\cos\phi}.
\end{equation}
\end{widetext}

Note that the above integrals are now written in terms of the coordinate system constructed in Eq \eqref{e:coord}, and illustrated in Figure 1, which was applied to \eqref{eq:HDsol} and \eqref{eq:kappa}.

Until now, one was only able to show this result by picking some values of pulsar distance 1, $L_1$,  and pulsar distance 2, $L_2$ and solve \eqref{eq1} numerically assuming some GW frequency $f$. In the literature, e.g. \cite{msmv13, ms14}, the authors invoke the reader's physical intuition to support the numerical result -- that if the exponents in \eqref{eq1} are large, $fL\gg1$, these oscillatory pieces rapidly converge to zero. This is often  referred to as the ``short wavelength approximation'', and has been used without proof, which we will now provide.

\subsection{Proof of Claim}

To prove this result, we estimate each of the four integrals \eqref{eq3}-\eqref{eq6} below which make up \eqref{eq1} separately. We apply the Lebesgue Dominated Convergence Theorem (see Appendix A), Fubini's Theorem, and the two-dimensional Divergence Theorem to get the required limiting value \eqref{eq2}. A key result used in the proofs which follow is a variant of the {\it Riemann-Lebesgue Lemma} in harmonic analysis (see \cite{bo}, p. 277 and \cite{sw}, p.2): let $a,b$ be finite (though this is not necessary). Then for a Lebesgue integrable function $f$ (a comprehensive definition and examples of this are given in Appendix A) over $[a, b]$, 
\beq
\label{eq:rlLemma}
\int_a^b dt \ e^{itx} f(t) \rightarrow 0 \hspace{0.5cm} \mathrm{as} \hspace{0.5cm} x\rightarrow \infty .
\eeq
The aforementioned Dominated Convergence Theorem, Equation \eqref{ldct}, basically gives us conditions under which we can interchange the operation of taking the limit of an integral with the integral of the limit.

First, we show that $K_2(\zeta,\phi,\theta)$ can be made continuous -- and so absolutely integrable over its domain of definition -- for all values of $ \theta \in [0,\pi]$, $\zeta\in (0, \pi]$, and $\phi \in [0,2\pi]$. 

We use the identity
\begin{eqnarray*}
&&1+\cos\theta\cos\zeta+\sin\theta\sin\zeta\cos\phi \\ 
&=& 1+\cos(\theta+\zeta)+\sin\theta\,\sin\zeta\,(1+\cos\phi),
\end{eqnarray*}
to show that the denominator of \eqref{k2}, i.e.,
$$1+\cos\theta\cos\zeta+\sin\theta\sin\zeta\cos\phi \geq 0$$
for all $ \theta \in [0,\pi]$, $\zeta\in (0, \pi]$, and $\phi \in [0,2\pi]$. It follows that the only singularities of $K_2$ must occur when the denominator vanishes, and this occurs precisely when $1+\cos(\theta+\zeta)=0$ and $\sin\theta\,\sin\zeta\,(1+\cos\phi)=0$, since both these quantities are necessarily non-negative. This, in turn, implies that for given $\zeta$, $\theta=\pi-\zeta$ and $\phi=\pi$ or $\zeta=0$, $\phi$ any, or $\zeta=\pi$, $\phi$ any. Each of these cases is handled by limiting arguments. 

For example, 
we note that
$$\lim_{\zeta\to 0^+} K_2(\zeta,\phi,\pi-\zeta) =0,\quad \lim_{\zeta\to \pi^-} K_2(\zeta,\phi,\pi-\zeta) =0,$$
and
$$\lim_{\phi\to \pi^-} K_2(\zeta,\phi,\pi-\zeta) = \frac{2\sin^3\zeta}{1-\cos\zeta}.$$
The previous equation gives a zero limit as $\zeta\to 0^+$ and is otherwise finite. It follows from this that $K_2$ can be defined to be a continuous function for any given value of $\zeta\in (0,\pi]$ and all values of $\theta \in [0,\pi]$ and $\phi \in [0,2\pi]$. Thus $K_2$ is Lebesgue integrable over the region $[0,2\pi]\times [0,\pi]$. Next,
\begin{widetext}
\begin{eqnarray}\label{eq3}
\int_0^\pi d\theta
\int_0^{2\pi}d\phi \,
 K_2(\zeta,\phi,\theta) \left[1- e^ {2\pi i fL_1(1+\cos\theta)} \right]\,\left[1-e^
{-2\pi i fL_2(1+\cos\theta\,\cos\zeta+\sin\theta\sin\zeta\cos\phi)} \right]\   =  
\end{eqnarray}
\begin{align}
& - \int_0^{\pi} d\theta \int_0^{2\pi} d\phi K_2(\zeta,\phi,\theta) \, e^ {2\pi {\rm i}   fL_1(1+\cos\theta)}
 \label{eq4x}\\
&- \int_0^{\pi} d\theta \int_0^{2\pi} d\phi K_2(\zeta,\phi,\theta) \, e^
{- 2\pi {\rm i} fL_2(1+\cos\theta\,\cos\zeta+\sin\theta\sin\zeta\cos\phi)} \ d\phi\, d\theta \label{eq4}\\
&+ \int_0^{\pi} d\theta \int_0^{2\pi} d\phi K_2(\zeta,\phi,\theta) \, e^
{2\pi {\rm i} \big (fL_1(1+\cos\theta) -
fL_2(1+\cos\theta\,\cos\zeta+\sin\theta\sin\zeta\cos\phi)\big)} \  \label{eq5}\\
&+ \int_0^{\pi}d\theta \int_0^{2\pi} d\phi K_2(\zeta,\phi,\theta) \label{eq6}
\end{align}
\end{widetext}
the last of which is identical to the required integral, \eqref{eq9}. Note that each of the previous four integrals is necessarily finite since the region of integration is finite and $K_2$ is absolutely integrable over it.

Write $\lambda:=fL_1, \mu :=fL_2$. Now we treat each of the previous three integrals \eqref{eq4x}-\eqref{eq5} separately, and fix $\zeta \in (0, \pi]$. 

\subsection*{Equation \eqref{eq4x} tends to zero}

Here we show that the first of the three equations with the exponential pulsar terms, \eqref{eq4x}, tends to zero. Using the above notation, and using the fact that $K_2$ is absolutely integrable over $\mathcal{R}$, Fubini's theorem on the interchange of iterated integrals yields the equality,
\begin{widetext}
$$I_1(\lambda):=  \int_0^{\pi} d\theta \int_0^{2\pi}d\phi\,  K_2(\zeta,\phi,\theta) \, e^ {2\pi {\rm i}  fL_1(1+\cos\theta)}
\   = \int_0^{2\pi} d\phi \, \left \{ \int_0^{\pi} d\theta\, K_2(\zeta,\phi,\theta) \, e^ {2\pi {\rm i} \lambda (1+ \cos\theta)}\right \},$$
\end{widetext}
which, after the change of variable $u=1+\cos\theta$, gives us,
$$I_1(\lambda) = \int_0^{2\pi} d\phi \, \left \{ \int_0^{2} du\, K_2^*(\zeta,\phi,u) \, e^ {2\pi {\rm i} \lambda u}\right \},$$
where $K_2^*(\zeta,\phi,u) = K_2(\zeta,\phi,\theta)/\sin\theta$ in the new variables is still absolutely integrable. Next, since $K_2^*$ is absolutely integrable over its domain, the ordinary two-dimensional version of the Riemann-Lebesgue Lemma, Equation \eqref{eq:rlLemma}, implies that

$$\lim_{\lambda\to\infty} \int_0^{2} du\, K_2^*(\zeta,\phi,u) \, e^ {2\pi {\rm i} \lambda u} = 0\, .$$

Since the previous integral is itself $\mathcal{O} (||K_2||_{\infty})$, the Lebesgue Dominated Convergence Theorem \eqref{ldct} can be used to interchange the order of the limit and the integral. We find:
$$\lim_{\lambda\to\infty} I_1(\lambda) = \int_0^{2\pi} d\phi \left \{ \lim_{\lambda\to\infty} \int_0^{2} du\, K_2^*(\zeta,\phi,u) \, e^ {2\pi {\rm i} \lambda u}\right \} = 0.$$
Thus, \eqref{eq4x} tends to zero as $\lambda \to \infty$. 

\subsection*{Equation \eqref{eq4} tends to zero }

{\bf Preamble} In order to extend the previous idea to more general exponents, we apply integration by parts to double integrals via the Divergence Theorem. In order to prove either \eqref{eq4} or \eqref{eq5} it suffices that we obtain the decay estimates $\mathcal{O}(1/\mu)$ or $\mathcal{O}(1/\lambda)$ as $\mu \to \infty$ or $\lambda \to \infty$. What follows is the general idea which we then apply to  the various cases. We need to estimate limits of the form
\begin{equation}\label{eq001}
I(\omega) := \iint_{\mathcal{D}} \,d\theta \,d\phi\,  f(\theta,\phi)\,  e^{i\omega\,g(\theta,\phi)} = \iint_ {\mathcal{D}} dA \, f e^{{\rm i}\omega g},
\end{equation}
as $\omega \to \infty$.  (Note that we used Fubini's Theorem to justify the interchange of the order of integration in the iterated integral \eqref{eq4}.) Here $\mathcal{D}$ along with its boundary (or perimeter), $\mathcal{C}$, are completely contained in $\mathcal{R}$ and are chosen so that $\nabla g(\theta,\phi) \neq \mathbf{0}$ on and inside $\mathcal{D}\cup \mathcal{C}$ (which necessarily has no points in common with $\mathcal{R}$).  By construction, the gradient of $g$, $\nabla{g}$ does not vanish on $\mathcal{D}\cup \mathcal{C}$ and therefore the quantity \begin{equation}\label{eq001a}\mathbf{u} = \frac{\nabla{g}}{|\nabla{g}|^2}\, f 
\end{equation}
 is well-defined on $\mathcal{D}\cup \mathcal{C}$.

We need to estimate the integral in \eqref{eq001} for large $\omega$. First, observe that (suppressing the variables for clarity of exposition)
$$\nabla \cdot (\mathbf{u}\, e^{{\rm i}\omega g}) = (\nabla\, e^{{\rm i}\omega g} )\cdot \mathbf{u}\, +  e^{{\rm i}\omega g}\, (\nabla\cdot \mathbf{u}),$$
where we assume, in addition, that $f$ is sufficiently smooth so that $\nabla\cdot \uu$ is defined. Since $\nabla g\cdot \mathbf{u} =  f$ we have
$\nabla (e^{{\rm i}\omega g})\cdot \mathbf{u} = {\rm i}\omega f e^{{\rm i}\omega g},$ which when inserted into the previous display and integrated over $\mathcal{D}$ yields, 
\begin{eqnarray*}
\iint_ {\mathcal{D}}  dA\,  \nabla \cdot (\mathbf{u}\, e^{{\rm i}\omega g})\ &= & \iint_ {\mathcal{D}} dA\, (\nabla \cdot \mathbf{u})\, e^{{\rm i}\omega g}+{\rm i}\omega \iint_ {\mathcal{D}}dA\, f e^{{\rm i}\omega g}\\
&=& \iint_ {\mathcal{D}} dA\, (\nabla \cdot \mathbf{u})\, e^{{\rm i}\omega g} +{\rm i}\omega I(\omega).
\end{eqnarray*}
 An application of the divergence theorem to the integral on the left gives us,
$$\iint_ {\mathcal{D}} dA\, \nabla \cdot (\mathbf{u}\, e^{{\rm i}\omega g})  = \int_{\mathcal{C}} d\sigma \, (\mathbf{u}\cdot \mathbf{n}) \,\, e^{{\rm i}\omega g}\,,$$
where $ \mathbf{n}$ is the unit normal to $\mathcal{C}$, itself oriented in the positive direction, and $\sigma$ is  arc length. Combining the two previous displays we get, 
\begin{equation}\label{002} 
I(\omega) = -\frac{{\rm i}}{\omega} \int_{\mathcal{C}} d\sigma \, (\mathbf{u}\cdot \mathbf{n})\, e^{{\rm i}\omega g} +\frac{{\rm i}}{\omega} \iint_{\mathcal{D}} dA \, (\nabla \cdot \mathbf{u})\, e^{{\rm i}\omega g}\, .
\end{equation}
Once we know that both integrands are absolutely integrable over $\mathcal{C}$ and $\mathcal{D}$ respectively, we get $I(\omega) = \mathcal{O} (1/\omega)$ or $I(\omega) \to 0$ as $\omega \to \infty$, over $\mathcal{D}\cup \mathcal{C}$. The results  \eqref{eq4} or \eqref{eq5} are obtained by a careful limiting analysis of the case where $\mathcal{D}\cup \mathcal{C}$ approaches $\mathcal{R}$ which then gives us the desired decay estimate.

{\bf Proof, Case 1:} $\zeta\in (0,\pi)$.  Set $g(\theta,\phi) = -(1+\cos\theta\,\cos\zeta+\sin\theta\sin\zeta\cos\phi)$, $\omega = 2\pi\,fL_2$ in \eqref{eq001}. Then $\nabla g (\theta,\phi)= \left (\sin\theta\sin\zeta\cos\phi,  \sin\theta\cos\zeta- \cos\theta\sin\zeta\cos\phi \right ),$ so that $\nabla g (\theta,\phi)=\mathbf{0}$ if and only if $\sin\theta\sin\phi=0$ and $\sin\theta\cos\zeta- \cos\theta\sin\zeta\cos\phi =0$. For $\zeta \in (0,\pi)$ this yields the eight (8) critical (or stationary) points 

\begin{eqnarray}
&& (\theta,\phi) = \left (0,\frac{\pi}{2}\right ), \, \left (0,\frac{3\pi}{2}\right ), \, \left (\pi,\frac{\pi}{2}\right ),\, \left (\pi,\frac{3\pi}{2}\right ), \nonumber \\
&& \left (\zeta, 0 \right), \left (\pi-\zeta, 0 \right ), \left (\zeta, 2\pi \right ), \left (\pi-\zeta, 2\pi\right ),
\end{eqnarray}
all of which are located on the perimeter of $\mathcal{R}$. Since we want $\mathcal{D}$ to be critical-point-free, for given $\varepsilon > 0$, choose $$\mathcal{D} = \{ (\theta,\phi) : \varepsilon< \theta < \pi - \varepsilon, \varepsilon < \phi < 2\pi - \varepsilon\},$$
and its perimeter,
$$\mathcal{C} = \{ (\theta,\phi) : \phi = \varepsilon,\, \phi = 2\pi - \varepsilon,\, \theta =\varepsilon,\, \theta=\pi - \varepsilon\}.$$
Then, by construction, $\nabla g \neq \mathbf{0}$ in $\mathcal{D}$ as well as on its perimeter, $\mathcal{C}.$ Defining $\uu $ as in \eqref{eq001a} we then obtain \eqref{002} for suitably smooth functions $f, g$, i.e., $I(\omega)\to 0$ on $\mathcal{D}\cup \mathcal{C}$, for every $\varepsilon > 0$. Now set $f=K_2$ and note that both integrals in \eqref{002} are finite on their respective region of integration.  Thus, for every $\varepsilon > 0$,
$$\lim_{fL_2\to \infty} \iint_{\mathcal{D}} \ d\theta\, d\phi K_2(\zeta,\phi,\theta) \, e^{{\rm i} \omega g(\theta,\phi)} =0, $$
i.e., so taking the limit as $\varepsilon$ approaches zero, we must have
\begin{equation}\label{eq001c}
\lim_{\varepsilon\to 0} \lim_{fL_2\to \infty} \iint_{\mathcal{D}} \ d\theta\, d\phi K_2(\zeta,\phi,\theta) \, e^{{\rm i} \omega g(\theta,\phi)} =0. 
\end{equation}
All that remains to be shown is that the interchange of the limits in the next expression is justified, i.e.,
\begin{widetext}
\begin{equation}\label{eq001b}
\lim_{\varepsilon\to 0} \lim_{fL_2\to \infty} \iint_{\mathcal{D}} \ d\theta\, d\phi \, K_2(\zeta,\phi,\theta) \, e^{{\rm i} \omega g(\theta,\phi)} =  \lim_{fL_2\to \infty} \lim_{\varepsilon\to 0} \iint_{\mathcal{D}} \ d\theta\, d\phi \, K_2(\zeta,\phi,\theta) \, e^{{\rm i} \omega g(\theta,\phi)},
\end{equation}
\end{widetext}
as the right hand side of \eqref{eq001b} is necessarily equal to \eqref{eq4} and so must  vanish as well by \eqref{eq001c}, which is what we set out to prove. To this end, we note that, by continuity of the integrals,

\begin{eqnarray}\label{eq001d}
\lim_{\varepsilon\to 0}&& \iint_{\mathcal{D}} d\theta\, d\phi \, K_2(\zeta,\phi,\theta) \, e^{{\rm i} \omega g(\theta,\phi)} \nonumber \\
&=&\iint_{\mathcal{R}} d\theta\, d\phi \,K_2(\zeta,\phi,\theta) \, e^{{\rm i} \omega g(\theta,\phi)}
\end{eqnarray}
and that, in fact, the convergence is uniform in $\omega$, for $\omega \in [0,\infty)$. 

\begin{widetext}
Indeed, observe that
\begin{eqnarray*}
\bigg | \iint_{\mathcal{R}}\ d\theta d\phi K_2(\zeta,\phi,\theta) e^{{\rm i} \omega g(\theta,\phi)} - \iint_{\mathcal{D}} \ d\theta d\phi K_2(\zeta,\phi,\theta) \, e^{{\rm i} \omega g(\theta,\phi)}\  \bigg | 
&=& \bigg | \iint_{\mathcal{R\setminus D}}\ d\theta d\phi K_2(\zeta,\phi,\theta) e^{{\rm i} \omega g(\theta,\phi)}\  \bigg | \\
&\leq&   \iint_{\mathcal{R\setminus D}} | \ d\theta\, d\phi\,K_2(\zeta,\phi,\theta)| 
\end{eqnarray*}
and this last integral may be made arbitrarily small, independently of $\omega$,  if $\varepsilon$ is sufficiently restricted. Hence the convergence in \eqref{eq001d} is uniform in $\omega$ (actually for any $\omega \in \R$ but we only require this for $\omega $ on the half axis, $[0,\infty)$). 
\end{widetext}

We are now in a position to apply a fundamental theorem on the interchange of such limits (see cite{pf}, p. 395)  to validate the equality in \eqref{eq001b} and complete the proof in the case where $\zeta \in (0, \pi).$

{\bf Case 2:} $\zeta=\pi$. In this case $g(\theta, \phi) = -1-\cos\theta$ is independent of $\phi$ and the resulting double integral can be handled in a similar way as \eqref{eq4x}, the only difference being the presence of a negative sign in the exponent. This, however, causes no difficulty with the argument in that section and so we omit the details.

\subsection*{Equation \eqref{eq5} tends to zero}

We use the same basic technique as in the proof of \eqref{eq4}. The proof of the limiting result for \eqref{eq5} can be obtained by reducing it to the case of \eqref{eq4} just proved. For example, \eqref{eq5} may be rewritten in the form
\begin{eqnarray*}
&&\int_0^{\pi} d\theta \int_0^{2\pi} d\phi \left \{ K_2(\zeta,\phi,\theta) \,
e^ {2\pi {\rm i }\lambda\, (1+\cos\theta)} \right\} \times \\
&&e^{- 2\pi {\rm i } \mu (1+\cos\theta\,\cos\zeta+\sin\theta\sin\zeta\cos\phi)} \ \ ,
\end{eqnarray*}
where now it is $K_2(\zeta,\phi,\theta) \, e^t
{2\pi {\rm i }\lambda\, (1+\cos\theta)}$ that is absolutely integrable over $\mathcal{R}$, since $K_2$ is and the exponential term has modulus equal to one. So, it follows from the methods above leading to \eqref{eq4} approaching zero as $\mu \to \infty$ that \eqref{eq5} also tends to zero as $\mu \to \infty$. Similarly, interchanging the $\mu$ and $\lambda$ terms in the preceding integral we obtain that \eqref{eq5} tends to zero as $\lambda \to \infty$ as well.

\subsection*{Final Result: The Hellings and Downs Curve}
We have shown that for an isotropic GWB, and for $fL_i \rightarrow \infty$, that the pulsar terms tend to zero. We can now write down the final form of the overlap reduction function:  the ``Hellings and Downs'' curve~\cite{hd83}:  
\begin{eqnarray}\label{eq9}
&& \int_0^\pi d\theta\, \int_0^{2\pi} d\phi\, K_2(\zeta,\phi,\theta)\, \\
&=&\frac{\sqrt{\pi}}{2}\! \bigg \{1 + \frac{\cos\zeta}{3} + 4(1-\cos \zeta)\ln \bigg (\frac{\sin\zeta}{2}\bigg ) \bigg \}(1+\delta_{1,2}), \nonumber 
\end{eqnarray}
for $\zeta \in (0, \pi]$ by \cite{hd83,abc+09, msmv13}.
Several comments should be made about Eq. \eqref{eq9} regarding a choice of normalization for the Hellings and Downs curve, the failure of the short-wavelength approximation, and the subsequent approximation of the pulsar term by a delta function for the autocorrelation.

When evaluating the autocorrelation, one can easily see that the value of the overlap reduction function is $4\sqrt{\pi}/3$. Indeed, it is a choice of normalization to set the autocorrelation equal to one when $\zeta=0$, which requires that Eq. \eqref{eq9} be multiplied by $3/(4\sqrt{\pi})$.

Next, one will note the $(1+\delta_{1,2})$ term: this takes into account the failure of the short-wavelength approximation when evaluating the autocorrelation term. We approximate this is a delta function, however, in Appendix C of \cite{grt+14}, Eq. C4, it is shown that the exact solution is
\beq
\Gamma_\mathrm{auto}(f,L_1=L_2) = \frac{1}{(2\pi f)^2}\bigg\{\frac{8\pi}{3}-\frac{1}{\pi(fL)^2} [1-j_0(4\pi fL)] \bigg\},
\eeq
where we adopt natural units ($c=1$). Here $j_0(x) =\sin x/x$ is a spherical Bessel function of the first kind. Since the oscillatory piece is suppressed by a factor of $1/(fL)^2$, approximating the pulsar term by a multiplicative factor of 2 for the $\zeta=0$ autocorrelation case is justified.

\citeauthor{ms14} 2014 \cite{ms14} also explored analytic expressions for the autocorrelation, and found that the $\kappa(fL_{1,2}, \hat\Omega)$ term, Eq. \eqref{eq:kappa}, can be well approximated as $2-2\cos M$, where $M=2\pi fL(1+\cos\theta)$. They showed numerically that the $2\cos M$ term contributed very little for large values of $fL$. 

We note that for small values of $\zeta$ and $fL$, there is an intermediate regime where one requires more than $(1+\delta_{1,2})$ to approximate $\kappa(fL_{1,2}, \hat\Omega)$. This case was explored in detail by \citeauthor{ms14} 2014 \cite{ms14}, who found that there are strong additional correlations from the pulsar term when the pulsars are separated by less than one gravitational wavelength. The authors also gave first and second order corrections for these cases.

\section{Discussion and Conclusion}
We have shown analytically that the Hellings and Downs curve approaches the Earth-term only solution, even when the pulsars are arbitrarily distant from the Earth, and not themselves at the same distance $L$ from the Earth. Of course, the case when $fL_1=fL_2:=fL$ is easily recovered, since in this case, $\lambda=\mu$ and all terms \eqref{eq4x}-\eqref{eq5} approach zero as the common value of this parameter $fL$ approaches infinity. 

The proofs indicate that the asymptotic estimate, Eq. \eqref{eq1}, holds for sufficiently smooth kernels that are absolutely integrable over the region $\mathcal{R}$, and not just kernels of the form $K_2(\zeta,\phi,\theta)$ as considered here. 

The astrophysical interpretation of this result is that if one monitors any galactic millisecond pulsar, and cross-correlates it with a pulsar in e.g. the Large Magellanic Cloud, the Hellings and Downs curve would still be correct correlation function to use, under the assumption that the GWB is isotropic. Anisotropic GWBs can be handled similarly, but care is required when evaluating the new kernel. 

To summarize, we have shown that for pulsars at distances $L_1$ and $L_2$ from the Earth, that the pulsar terms tend to zero as the $fL_i\to \infty$. The asymptotic estimate \eqref{eq2} is false if $fL_2$ is fixed as there is no reason for the integral \eqref{eq4} to tend to zero as $fL_1\to \infty$, since it is independent of $fL_1$. 
While this result is consistent with the previous intuition developed in the field of nanohertz GW astronomy, and indeed verified numerically for a few values of $fL$ in \cite{ms14}, it has never before been proven analytically or generally for any $fL_i$. This result is an important validation of a fundamental result in the field, and lends credibility and rigor to current GWB searches as we enter detection era in nanohertz GW astronomy.

\acknowledgements
The authors thank Yacine Ali-Ha{\"i}moud for useful comments and a thorough reading of this manuscript. Figure \ref{fig:gwspectrum} was generated in part with the online tool ``gwplotter''~\cite{mcb15}.  The Flatiron Institute is supported by the Simons Foundation.

\appendix 
\section{Overview of Lebesgue Integration}

Here we give a brief overview of the {\it Lebesgue integral} on the real line, $\R$, see \citep{wr}, Chapter 10.
Readers familiar with this concept may proceed immediately to the Main Results in section III and their proofs.

One of the great advantages of the Lebesgue integral over the classical (Riemann) integral is in the handling of limiting processes such as limits of integrals of sequences of functions which, in the classical case, usually requires uniform convergence of the sequence in question -- this is not the case for the Lebesgue integral. 

Fundamental to this now-standard integral in mathematics is the notion of {\it Lebesgue measure}. One intuitive notion of measure assigns the value $b-a$ for the length of an interval $[a,b]$ or, more generally, $(b-a)(d-c)$, for the area of a plane rectangle $\mathcal{R}=[a,b]\times [c,d] =\{(x,y) : x \in [a,b]\ \ {\rm and}\ \ y\in [c,d]\} $ formed by the Cartesian product of the two intervals $[a,b]$ and $[c,d]$. The notion of measure allows one to extend this property (length, area, volume, etc) to Lebesgue measurable sets. Aside traditional examples such as rectangles, unions of disjoint intervals, etc. one can consider the set $\Q$ of rational numbers on the real line, or its complement, the set of irrational numbers there. Both the latter two sets are Lebesgue measurable, the former has Lebesgue measure zero while the latter has positive Lebesgue measure. 

More precisely, a set $E \subset \R$ is said to have Lebesgue {\it measure zero} if for any given $\varepsilon > 0$ there is a sequence of intervals $[a_n, b_n]$, n= 1,2,\ldots  such that $E$ is contained within the union of all these intervals where, in addition, $\sum_{n=1}^\infty (b_n-a_n) < \varepsilon$. For example, the set of all rational numbers $\Q \subset \R$ has measure zero.

One of the connections between the Lebesgue theory and the ordinary Riemann theory of the integral is the following result: {\it If a function $f$ is bounded on an interval $[a,b]$ then $f$ is Riemann integrable over $[a,b]$ if and only if the set of its discontinuities is a set of measure zero.}  By its very definition, the Lebesgue integrability of $f$ forces that the absolute value of the function, $|f|$, in question be Lebesgue integrable, which is not so in the case of the Riemann integral. Still, the advantage in using Lebesgue integrals is huge in that we can extend the class of functions and sets over which we are integrating and still get meaningful results.

Now we offer an extremely brief introduction to the Lebesgue integral, with a few key definitions and results. Briefly, with the Lebesgue integral we subdivide the {\it range} of a given function $f$, look for those parts where horizontal lines intersect the graph of $f$ and then drop rectangles onto the domain of $f$. (Recall that we subdivide the {\it domain} of $f$ in the case of the Riemann integral.)  

More generally, the measure of a set $E \subset \R$ is the greatest lower bound (see e.g. \cite{wr}, p. 11) of the set of all numbers of the form $\sum_{n=1}^\infty (b_n-a_n)$ where the union of all the intervals $[a_n, b_n]$ contains $E$. A set $E$ is said to be {\it Lebesgue measurable} if it has finite measure. A function $f$ is said to be measurable if the special set $F=\{ x : f(x) > a\}$ is measurable for every real number $a$. These are precisely the functions that we can integrate in the Lebesgue sense, i.e., the Lebesgue integrable functions over $E$. First, the integral is defined for {\it simple functions} (i.e., those functions whose range is a finite set of points). Then, using the fact that for a given measurable function $f(x) \geq 0$ there is a monotonically increasing sequence of simple functions $s_n(x)$ that converges to $f(x)$ and whose integrals, $\int_{E} s_n(x)\, dx$, are bounded by a constant $C$ (that depends on $f$), we define the Lebesgue integral of $f$ over $E$ by
$$\int_{E} f(x)\, dx = \lim_{n\to \infty} \int_{E} s_n(x)\, dx,$$
Finally, when $f$ is a general measurable function (i.e., not necessarily non-negative) we define its integral using its  decomposition into positive and negative parts, that is, we know that $f(x) = f^+(x) - f^-(x)$ where $f^{\pm}(x) = \max \{\pm f(x), 0\}$. Thus, the Lebesgue integral of a general measurable function $f$ over $E$ is by definition,
$$\int_{E} f(x)\, dx = \int_{E} f^+(x)\, dx  - \int_{E} f^-(x)\, dx,$$
It is then an easy matter to see that
$$\int_{E} |f(x)|\, dx = \int_{E} f^+(x)\, dx  + \int_{E} f^-(x)\, dx.$$
The Lebesgue Dominated Convergence Theorem states that if $f_n(x)$ is a sequence of measurable functions with $f_n(x) \to f(x)$ almost everywhere on $E$. (This means that $f_n(x) \to f(x)$ at all points $x$ except for a set of measure zero.) In addition, let  $g\geq 0$ be Lebesgue integrable on $E$ with $|f_n(x) | \leq g(x)$ then
\begin{equation}\label{ldct}
\lim_{n\to \infty} \int_{E} f_n(x)\, dx = \int_{E} f(x)\, dx.
\end{equation}
This theorem is the main one being used in this paper in order to guarantee convergence of the various integrals. A similar theory and similar results hold in 2 or more dimensions.

The space  $L^1[0,\pi]$ is by definition the space of all complex valued integrable functions $f$ such that $$\int_0^{\pi} |f(x)|\, dx < \infty.$$ In addition, the space $L^\infty(\R^+)$ is the space of all such functions $f$ for which there exists a constant, $C$, depending on $f$, such that 
$|f(x)| < C$ almost everywhere on $\R^+$.


\end{document}